\documentclass{ws-ijmpe}

\begin{document}

\markboth{Upadhyay, Kelkar, Khemchandani, Jain}{Study of the $\eta-^7Be$ 
interaction in $p-^6Li$ fusion}

\catchline{}{}{}{}{}
%%%%%%%%%%%%%%%%%%%%%%%%%%%%%%%%%%%%%%%%%%%%%%%%%%%%%%%%%%%%%%%%%%%%

\title{STUDY OF THE $\eta-^7Be$ INTERACTION NEAR THRESHOLD IN $p-^6Li$
FUSION}

\author{N. J. UPADHYAY,$^{1}$\footnote{njupadhyay@gmail.com}  
N. G. KELKAR,$^2$ K. P. KHEMCHANDANI,$^3$ B. K. JAIN$^1$
%N. G. KELKAR$^{2\,,}$\footnote{nkelkar@uniandes.edu.co}  , 
%K. P. KHEMCHANDANI$^{3\,,}$\footnote{kanchan@ific.uv.es}  , 
%B. K. JAIN$^{1\,,}$\footnote{brajeshk@gmail.com}
}

\address{$^1$Department of Physics, University of Mumbai,
Mumbai - 400 098, INDIA.
\\
$^2$Department de Fisica, Universidad de los Andes,
Bogota, COLOMBIA.
\\
$^3$Departamento de F\'isica Te\'orica $\&$ IFIC,
Universidad de Valencia-CSIC, Valencia, SPAIN.}
\maketitle

\begin{history}
\received{(received date)}
\revised{(revised date)}
%\accepted{(Day Month Year)}
%\comby{(xxxxxxxxxx)}
\end{history}

\begin{abstract}
We present a calculation for $\eta$ production in the $p-^6Li$ fusion
near threshold including the $\eta-^7Be$ final state interaction
(FSI). We consider the $^6Li$ and $^7Be$ nuclei as $\alpha-d$ and
$\alpha-^3He$ clusters respectively. The calculations are done for
the lowest states of $^7Be$ with $J\,=\,\left(\,{3\over2}^-\,,\,
{1\over2}^-\,\right)$ resulting from the $L\,=\,$1 radial wave function.
The $\eta-^7Be$ interaction is incorporated through the $\eta-^7Be$ 
$T-$matrix, constructed from the medium modified matrices for the 
$\eta-^3He$ and $\eta-\alpha$ systems. These medium modified matrices 
are obtained by solving few body equations, where the scattering in 
nuclear medium is taken into account.
\end{abstract}

\section{Introduction}

The search for exotic $\eta-$nucleus quasi-bound states \cite{haili} 
is guided by the strong and attractive nature of the $\eta-N$ 
interaction in the $s-$wave \cite{bhale}.
With this motivation, measurements for the $^6Li\,(\,p\,,\,\eta\,)\,^7Be$
reaction \cite{scomp} were carried out by the Turin group in 1993
at an incident energy of 683 MeV. A theoretical
study of this reaction \cite{khal} was made and it was summarized that 
the information
available was not sufficient to conclude the formation of an $\eta-^7Be$ 
quasi-bound state. The interest in this reaction has been revived by the 
recent sophisticated study of this reaction at the COSY Laboratory at an
incident energy of 673 MeV \cite{mach}. Using the experience gained by 
studying the interaction of $\eta$ with lighter nuclei like the deuteron 
\cite{bkj} and $^3He$ \cite{kan} by solving few body equations, we extend 
our study to a heavier nucleus, {\it i.e.}, the $^7Be$.

\section{The Formalism}
\begin{figure}[th]
\centerline{\psfig{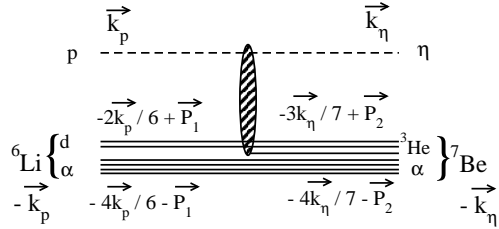}}
\vspace*{8pt}
\caption{The reaction mechanism for $p\,^6Li\,\rightarrow\,\eta\,^7Be$ 
reaction.}
\end{figure}

\subsection{The production mechanism}

We consider $^6Li$ and $^7Be$ nuclei as $\alpha-d$ and $\alpha-^3He$
clusters. The reaction mechanism is, hence, built by assuming a 
collision of the beam proton with 
the deuteron in $^6Li$ producing an $\eta$ and $^3He$. Combining with 
the spectator $\alpha$, $^3He$ produces $^7Be$ in a particular state 
(see Fig. 1). The production $T-$matrix for a relative angular momentum 
$L$ between the $^3He$ and $\alpha$ is written as
\begin{eqnarray}
\nonumber
\langle\,\vec{q}\,|\,T_{p\,^6Li\,\rightarrow\,\eta\,^7Be}\,|\,\vec{k_p}\,
\rangle&=&i^{(L\,+\,1)}\,\sqrt{4\,\pi}\,\sum_{M\,\mu}Y^*_{L\,M}(\hat{Q})\,
F_L(Q)\,\langle\,{1\over2}\,,\,L\,;\,\mu\,,\,M_L\,|\,J\,,\,m_7\,\rangle
\\
&&\times\,\langle\,{1\over2}\,,\,\mu\,|\,T_{p\,d\,\rightarrow\,\eta\,^3He}\,
|\,{1\over2}\,,\,m_p\,;\,1\,,\,m_6\,\rangle
\end{eqnarray}
where, $F_L(Q)\,=\,\int_0^\infty\,r^2\,dr\,(\,\psi^{*\,^7Be}_L(r)\,j_L(Q\,r)\,
\psi^{^6Li}_0(r)\,)$ is the transition form factor for $^6Li\,\rightarrow\,
^7Be$, with $\vec{Q}\,=\,{4\over7}\vec{q}\,-\,{2\over3}\vec{k_p}$. Here,
$\psi^{^6Li}_0(r)$ and $\psi^{*\,^7Be}_L(r)$ are the radial wave functions for
$^6Li$ and $^7Be$ respectively. The $T-$matrix for the elementary process, 
$\langle\,|\,T_{p\,d\,\rightarrow\,\eta\,^3He}\,|\,\rangle$, 
is written in a two-step model from our earlier work \cite{kan}. 
The calculations have been carried out by neglecting the effect of Fermi 
motion on $\langle\,|\,T_{p\,d\,\rightarrow\,\eta\,^3He}\,|\,\rangle$.

The transition matrix for the $p\,^6Li\,\rightarrow\,\eta\,^7Be$ reaction,
including the $\eta-^7Be$ FSI is given as
\begin{eqnarray}
T&=&\langle\,\vec{k_{\eta}}\,,\,m_7\,|\,T_{p\,^6Li\,\rightarrow\,\eta\,^7Be}
\,|\,\vec{k_p}\,;\,m_p\,,\,m_6\rangle\\
\nonumber
&&+\,\sum_{m_7^\prime}\int {d\vec{q}\over(2\,\pi)^3}\,{\langle\,\vec{k_{\eta}}
\,,\,m_7\,|\,T_{\eta\,^7Be}\,|\,\vec{q}\,;\,m_p\,,\,m_7^\prime\,\rangle \over 
E(k_{\eta})\,-\,E(q)\,+\,i\,\epsilon}\,\langle\,\vec{q}\,,\,m_7^\prime\,|\,
T_{p\,^6Li\,\rightarrow\,\eta\,^7Be}\,|\,\vec{k_p}\,;\,m_p\,,\,m_6\rangle
\end{eqnarray}
where $\vec{k_p}$ and $\vec{k_{\eta}}$ are the initial and final momenta in 
the centre of mass system. $m_p$, $m_6$ and $m_7$ are the spin projections for 
the proton, $^6Li$ and $^7Be$ respectively.

\subsection{Final state interaction}

The $\eta-^7Be$ interaction is incorporated through a half-off-shell 
$\eta-^7Be$ $T-$matrix,
\begin{equation}
T_{\eta^7Be}(k^\prime\,k\,z)=\int d\vec{x_1}\,
|\psi_{\alpha\,^3He}(x_1)|^2\,\{T_3(k^\prime\,,\,k\,,\,a_1x_1\,,\,z)
+T_{\alpha}(k^\prime\,,\,k\,,\,a_2x_1\,,\,z)\}
\end{equation}
where, $T_3(k^\prime\,,\,k\,,\,a_1x_1\,,\,z)$ and 
$T_{\alpha}(k^\prime\,,\,k\,,\,a_2x_1\,,\,z)$ represent $\eta-^3He$ and 
$\eta-^4He$ scattering matrices, respectively. They are written as,
\begin{eqnarray}
\nonumber
T_3(k^\prime\,,\,k\,,\,a_1x_1\,,\,z)\,=\,t_3(k^\prime\,,\,k\,,\,a_1x_1\,,\,z)
&+&{1 \over 2\pi^2}\int_0^\infty q^2\,dq\,
{t_3(k^\prime\,,\,q\,,\,a_1x_1\,,\,z) \over \left(\,z\,-\,{q^2\over2\mu}\,
\right)}
\\
&&\times \,T_{\alpha}(k^\prime\,,\,k\,,\,a_2x_1\,,\,z)
\\
\nonumber
T_{\alpha}(k^\prime\,,\,k\,,\,a_2x_1\,,\,z)\,=\,t_{\alpha}
(k^\prime\,,\,k\,,\,a_2x_1\,,\,z)&+&{1 \over 2\pi^2}
\int_0^\infty q^2\,dq\,
{t_{\alpha}(k^\prime\,,\,q\,,\,a_2x_1\,,\,z) \over \left(\,z\,-\,
{q^2\over2\mu}\,\right)}
\\
&&\times \,T_3(k^\prime\,,\,k\,,\,a_1x_1\,,\,z)
\end{eqnarray}
where, $x_1$, the initial Jacobi co-ordinate, is related to the  
position vector, $\vec{r_i}$ 
in the $^3He\,-\,\alpha\,-\,\eta$ centre of mass system of each 
of the constituents by
$\vec{r_i}\,=\,a_i\,\vec{x_1}$. $z\,=\,
E\,-\,|\epsilon_0|\,+\,i\,0$ with $|\epsilon_0|$ being the energy 
required for the break up of $^7Be$ into $^3He$ and $\alpha$.
The $t_3(k^\prime\,,\,k\,,\,a_1x_1\,,\,z)$ and 
$t_{\alpha}(k^\prime\,,\,k\,,\,a_2x_1\,,\,z)$ matrices have been 
calculated as in Ref. [7].

To begin with, we assume that the $\eta$-meson scatters only once 
from each of the
$^7Be$ constituents, {\it i.e.}, $^3He$ and $\alpha$ . 
The required single scattering terms, the
 $t_3(k^\prime\,,\,k\,,\,a_1x_1\,,\,z)$ 
and $t_{\alpha}(k^\prime\,,\,k\,,\,a_2x_1\,,\,z)$ in Eq. (4) and 
Eq. (5) are 
constructed using the factorized impulse approximation (FIA) 
\cite{cres,khal2}. In this approximation, the momentum 
in the $\eta\,^3He$ and $\eta\,\alpha$ centre of masses are 
obtained by boosting their momenta in the $\eta-^7Be$ 
centre of mass. This includes the Fermi momenta of 
$^3He$ and $\alpha$ in $^7Be$. 

\subsection{$^6Li$ and $^7Be$ wave functions}

We have used two prescriptions for $\psi_0^{^6Li}$ and $\psi_L^{*\,^7Be}$ 
cluster wave functions, namely (1) Cluster model wave function generated using 
Wood-Saxon potential \cite{khal,buc,kuk} and (2) Green's function Monte Carlo
(GFMC) Variational wave function generated using Urbana potential \cite{for}.

Since the energy spacing between the first four low-lying levels of the 
$^7Be$ is very small, it is important to include the contribution from these
levels. In the present calculation we consider the lowest states of $^7Be$,
$\left(\,{3\over2}^-\,,\,{1\over2}^-\right)$ doublet split by 0.48 MeV, 
resulting from the $L\,=\,1$ radial wave function coupled to spin ${1\over2}$ 
of the $^3He$.

\section{Results and Discussion}

In Fig. 2, we show the total cross sections as a function of the excess 
energy. In this figure we show the plane wave results and those obtained 
after including the $\eta-^7Be$ FSI. 
We find that effect of inclusion of the FSI changes the shape and the 
magnitude of the total cross section curve drastically. In this
plot we also show sensitivity of the results to different choice of
wave functions. The FSI matrix has been calculated using an $\eta N$ 
$t-$matrix which corresponds to $a_{\eta N}\,=\,0.88 + i0.41$ fm. 
\begin{figure}[th]
\centerline{\psfig{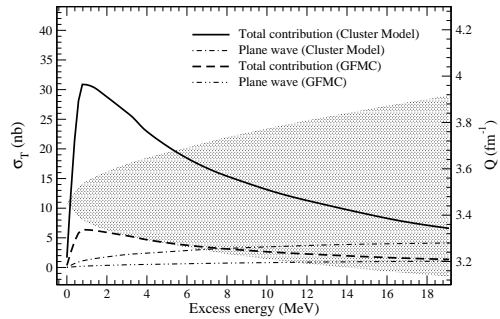}}
\vspace*{8pt}
\caption{Total cross section as a function of Q for 
$a_{\eta\,N}$ = 0.88 + i0.41 fm. The results are shown for two different
prescriptions of cluster wave functions using the FIA approach.}
\end{figure}
\begin{figure}[th]
\centerline{\psfig{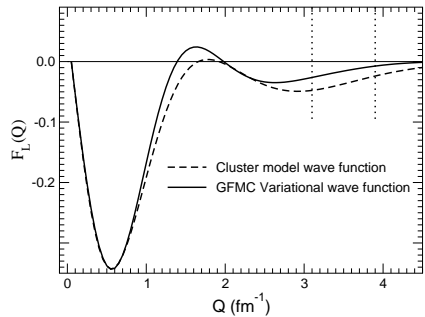}}
\vspace*{8pt}
\caption{Form factor, $F_L(Q)$ as a function of $Q$ for $L\,=\,1$ state for
two different prescriptions of cluster wave functions.}
\end{figure}

In Fig. 3 we show the form factor, $F_L(Q)$, calculated using two different 
models. The region between the dotted lines corresponds to the range of $Q$, 
at which $F_L(Q)$ in Eq. (1) is getting calculated. This range of $Q$ is 
shown in Fig. 2 by hashed region. It can be seen that two form factors differ 
in this region which gives rise to the difference in the total cross sections 
calculated using the two different wave functions.

\begin{figure}[th]
\centerline{\psfig{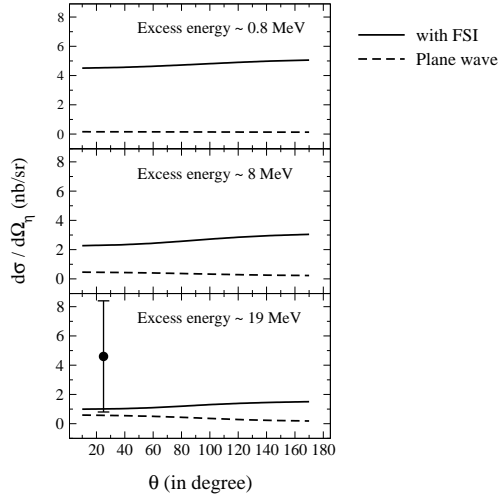}}
\vspace*{8pt}
\caption{The calculated angular distribution for cluster model wave function
using the FIA approach.}
\end{figure}

In Fig. 4, we show the angular distributions for three different
excess energies such that it spans region from close to threshold up to 
20 MeV above threshold. We show results without FSI and with FSI for 
cluster model wave functions. We find that even after including FSI, 
the angular distribution is isotropic in nature. At excess energy 
$\sim\,$19 MeV, the differential cross section is $\sim\,$1 nb/sr
for low-lying states of $^7Be$ resulting from the $L\,=\,1$ radial wave 
function, in comparison with the experimental number \cite{scomp}, 
{\it i.e.}, $4.6 \,\pm\,3.8$ nb/sr.  

A calculation of contributions from the excited states of the $^7Be$ nucleus
to the total cross section and that of more detailed FSI is in progress and 
shall soon be reported.

\end{document}